\def\gtorder{\mathrel{\raise.3ex\hbox{$>$}\mkern-14mu
	     \lower0.6ex\hbox{$\sim$}}}
\def\ltorder{\mathrel{\raise.3ex\hbox{$<$}\mkern-14mu
	     \lower0.6ex\hbox{$\sim$}}}

\def\eqsim{\mathrel{\raise.4ex\hbox{$=$}\mkern-14mu
	     \lower0.6ex\hbox{$\sim$}}}

\documentstyle[preprint,prc,aps]{revtex}

\draft
\begin{document}

\title{Nuclear correlation and finite interaction-range effects \\
       in high-energy $(e,e'p)$ nuclear transparency}

\author{Ryoichi Seki${}^{1,3}$, T. D. Shoppa${}^1$,
	Akihisa Kohama${}^2$, and Koichi Yazaki${}^2$}

\address{ ${}^1$ W. K. Kellogg Radiation Laboratory,
	California Institute of Technology, Pasadena,
	CA 91125}
\address{ ${}^2$ Department of Physics, Faculty of Science,
	University of Tokyo,
	7-3-1 Hongo Bunkyo-ku Tokyo 113, JAPAN}
\address{ ${}^3$ Department of Physics and Astronomy,
       California State University, Northridge, CA 91330}

\date{\today}
\maketitle

\begin {abstract}

Nuclear transparency is calculated for high-energy, semi-inclusive
$(e,e'p)$ reactions, by accounting for all orders of Glauber
multiple-scattering and by using realistic finite-range $p N$ interaction
and (dynamically and statistically) correlated nuclear wave functions.
The nuclear correlation effect is reduced due to the $p N$ finite-range
effect.  The net effect is small, and depends sensitively
on details of the nuclear correlations in finite nuclei, which are
poorly known at present.

\end{abstract}


\narrowtext

Perturbative quantum chromodynamics (pQCD) predicts
the novel nuclear phenomenon
of color transparency, in which a hadron produced by a hard
scattering process could have an unusually long mean-free-path
in nuclei~\cite{Bro:Mue}.
Such a hadron would be small, color neutral, and suffer little interaction
with other nucleons as it goes through the nucleus.
The phenomenon has been observed
in $(p, 2p)$~\cite{Car:PRL} and vector meson production~\cite{Ada:PRL},
while it is not seen in $(e,e'p)$~\cite{NE18}.  pQCD calculations
largely remain semi-phenomenological~\cite{INTER:CEBAF} and
have not yet yielded precise quantitative predictions.

Low-energy hadrons can have long mean-free-paths in nuclei solely due to
nuclear medium effects without color transparency~\cite{PandhaPieper}.
It is imperative to establish nuclear medium effects reliably
in these high-energy reactions,
so as to identify the pQCD phenomenon explicitly.
The medium effect that has been most noted is nuclear correlation,
but calculations vary wildly
from only a few percent increase~\cite{Nik:PL} to as much as a 20
to 30 percent increase~\cite{Ben:PRL} in proton emission.

The correlation effect results from an interplay
between the nuclear reaction mechanism and nuclear structure.
As such, both must be formulated and computed with equal rigor and
care in order to unambiguously establish the significance of
the effect.
In this letter we report results of calculations of nuclear transparency
in $(e,e'p)$.
We include all orders of Glauber multiple-scattering series with
realistic finite-range $p N$ interaction together with realistic
nuclear correlations, and carry out numerical evaluation using
Monte Carlo integration over all nucleon coordinates and spins.
No previous work has been carried out with this rigor~\cite{Frankel}.

The central quantity in this work is the nuclear
transparency $T$, which corresponds to a ratio of the proton emission
rate with and without the final-state interaction.
$T$ is given as the missing-momentum integration of the response
function,
\begin{eqnarray}
S^{\uparrow}  = -{1 \over \pi}
< 0 | O^{\dagger} \cdot Im G^{\uparrow} \cdot O | 0 >
\label{eq:response}
\end{eqnarray}
for the ground-state nucleus $| 0>$.  Here, $Im G^{\uparrow}$ is
the imaginary part of the Green's function due to proton emission, and
\begin{eqnarray}
Im G^{\uparrow} = (1 + G^{\dagger} V^{\dagger}_{1,A-1})
		   Im G_0 (1 + V_{1,A-1} G),
\label{eq:green}
\end{eqnarray}
where $G_0$ and $G$ are the Green's function for the proton in free space
and in the nucleus, and
$V_{1,A-1}$ is the interaction between the knockout high-momentum
proton (labeled as 1) and the rest of the $A-1$ nucleons~\cite{Mori:Yaza}.

We apply
the Glauber approximation for Green's functions, which consists of
the eikonal approximation for the proton and the fixed-scatterer
approximation for the target nucleons, and obtain
\begin{eqnarray}
 T = {\int d {\bf r}_1 \cdots d {\bf r}_A
 {\Pi}_{j=2}^A e^{ - 2 Im \chi_j ( {\bf r}_1, {\bf r}_j ) }
 {|\Psi ({\bf r}_1,\cdots,{\bf r}_A)|}^2 \over
 \int d {\bf r}_1 \cdots d {\bf r}_A
 {|\Psi ({\bf r}_1,\cdots,{\bf r}_A)|}^2}
\label{eq:tran2}
\end{eqnarray}
with
\begin{eqnarray}
{\chi}_j ({\bf r}_1 , {\bf r}_j) =
-{ m \over p_1} \int_{z_1}^{\infty} d {z_1}'
V_{pN} { ({\bf r}_1^{'} - {\bf r}_j ) }_{{\bf b}'_1 = {\bf b}_1}.
\label{eq:opr1}
\end{eqnarray}
Equation (\ref{eq:tran2}) appeared in our previous work~\cite{Koh:NP}
and and also in Ref.~\cite{Nik:PL} as an intermediate step
to Glauber multiple-scattering series~\cite{Glo}.

$T$ can be cast in the well-known eikonal form expressed
in terms of the one-body nuclear density, $\rho ({\bf r})$:
\begin{eqnarray}
T ={
\int d {\bf r}_1  exp ( - 2 Im \chi({\bf r}_1))
\rho ( {\bf r}_1)
\over
\int d {\bf r}_1 \rho ({\bf r}_1) }.
\label{eq:tran3}
\end{eqnarray}
Here, 2$Im \chi( {\bf r}_1 )$ is expressed in a cluster expansion as
\begin{eqnarray}
{\sum}_{j=2}^A \int d {\bf r}_j
(2 Re \Gamma_j -|\Gamma_j|^2)
\rho_2 ( {\bf r}_1,{\bf r}_j ) / \rho ( {\bf r}_1 ) + \cdots,
\label{eq:tran4}
\end{eqnarray}
where $\Gamma_j =1 - exp(+i{\chi}_j ({\bf r}_1 , {\bf r}_j))$
is the Glauber $N N$ amplitude in the coordinate space.

Nuclear correlation effects appear explicitly in the leading order
through the two-body nuclear density $\rho_2$, and spectator effects
appear in the subleading order (not explicitly shown here)~\cite{Nik:PL}.
The commonly used eikonal expression is a consequence of
invoking the zero-range approximation and setting
$\Gamma_j$ as a constant proportional to the $N N$ cross section, $\sigma$:
Either
\begin{eqnarray}
2 Im \chi( {\bf r}_1 ) \approx
 \sigma \int_{z_1}^{\infty} dz'_1
  g( {\bf r}_1 - {\bf r'}_1){\rho ({\bf r'}_1)}_{{\bf b'}_1 = {\bf b}_1}
\label{eq:tran5}
\end{eqnarray}
or
\begin{eqnarray}
2 Im \chi( {\bf r}_1 ) \approx
 \sigma \int_{z_1}^{\infty} dz'_1
 {\rho ({\bf r'}_1)}_{{\bf b'}_1 = {\bf b}_1}
\label{eq:tran6}
\end{eqnarray}
with or without the correlation, repectively.
Furthermore, many authors use the total cross section, $\sigma_{tot}$,
as $\sigma$.  This corresponds to discarding the $|\Gamma_j|^2$
term in Eq.~(\ref{eq:tran4}) and is not justified
for a fully semi-inclusive process~\cite{Koh:NP} \cite{Rinat}
since the term is of the leading order through
unitarity~\cite{Glo},~\cite{Koh:NP}.
Note that the use of $\sigma_{tot}$ is also inappropriate in exclusive
(e,e'p) reactions~\cite{Moniz}.

$V_{pN}$ in Eq.~(\ref{eq:opr1}) is either the $pp$ or $pn$ potential,
depending on whether the $j$-th nucleon is the proton or neutron.
We determine $V_{pN}$ by applying the Abel transformation
to invert the eikonal forms of the $N N$ amplitudes~\cite{Glo}.
The inversion is exact in the eikonal formalism for a local,
spherical potential; we thus determine $V_{pN}$ better than
the frequently used Born approximation.
The $N N$ amplitudes are extracted from experimental data
in Gaussian forms~\cite{Tom}.
Because of the limited availability of $p N$ scattering data,
we also restrict the potentials to be spin-independent.
The free-space $p N$ interaction would be modified
by nuclear medium corrections and by long formation lengths of
excited nucleons.  These are clearly of much interest, but we proceed
by neglecting them in order to determine
the correlation effects with minimal complication.

We now describe our findings by taking the case of
$p_1 = 4.49$ GeV/c for doubly closed shell nuclei, ${}^{16}$O and ${}^{40}$Ca.
This choice is also made to minimize effects of other matters that might
complicate the issue, such as open nuclear shells and strong
momentum dependence of the $p N$ amplitudes.
Note that the $p_1$ value corresponds to the highest $Q^2$ in the NE18
$(e,e'p)$ experiment~\cite{NE18}.

As Eq.~(\ref{eq:tran2}) shows, $exp(-2 Im \chi ({\bf r}, 0))$ contributes
to $T$ as a product with the correlated nuclear density.
In Fig.~1, we illustrate
$exp(-2 Im \chi ({\bf r}, 0))$ calculated from Eq.~(\ref{eq:opr1})
using the $V_{pp}$ inverted from the observed $p p$ amplitude.
 Figure 1 shows
that $exp(-2 Im \chi ({\bf r},0))$ deviates from unity even for $b$ and
$z > 1$ fm.  This feature is missing in the commonly
used approximation of the zero-range $p N$ interaction:
The corresponding figure in the zero-range approximation would appear
as a deep narrow valley following the z-axis starting
from $z = 0$.

In order to evaluate $T$, we require
the $A$-body nuclear density that is accurate in the length scale
of several tenths of fm or smaller.  Since such densities are not readily
available for various nuclei, we construct them by following
a simple scheme and compare with a variational Monte Carlo calculation of
${}^{16}$O~\cite{PWP:PRC}.  We use a Jastrow-form of the wave function:
a product of antisymmetrized Hartree-Fock (HF) single-particle wave function
(for Skyrme force~\cite{Rein:Comp}) ${\it A} \Phi$
and of pair wave functions
for nuclear matter at various densities~\cite{Wir:Pie} $f$'s,
which describe the statistical and dynamical correlations, repectively.
The A-body density $|\Psi ({\bf r}_1,\cdots,{\bf r}_A)|^2$ is
\begin{eqnarray}
& & |{\it A} \Phi|^2_s \cdot
\prod_{i < j}
 [ ( {f_c}^2 + 3 {f_{\sigma}}^2 + 3 {f_{\tau}}^2 + 9 {f_{\sigma \tau}}^2
     +6{f_t}^2 + 18 {f_{t \tau}}^2) \nonumber \\
& & \pm 2 ( f_c f_{\tau}
 + 3 f_{\sigma} f_{\sigma \tau} + 6 f_t f_{t \tau}
  - {f_{\tau}}^2 - 3 {f_{\sigma\tau}}^2 -6 {f_{t \tau}}^2    )]
\label{eq:dens}
\end{eqnarray}
for a $p n$ pair $(-)$ and a $p p$ or $n n$ pair $(+)$.
Invoking the local density approximation, we evaluate the $f$'s
as functions of $|{\bf r}_i - {\bf r}_j |$,
for the (one-body) nuclear density
that is generated from $|{\it A} \Phi|^2_s$ (spin-summed)
at $|{\bf r}_i + {\bf r}_j|/2$.

Here, we include the dynamical correlations only between nucleon pairs,
neglecting correlation operator symmetrization and
off-diagonal contributions.
The terms depending linearly on the spin-dependent operators are discarded,
and those depending quadratically are averaged.
This simple scheme should be reasonable for closed-shell nuclei,
especially because all noncentral $f$'s
are much smaller than the central $f_c$.  The tensor correlations
become stronger relative to $f_c$ at short distances ($\ltorder$ 0.5 fm),
affecting significantly the momentum distribution above the Fermi
momentum~\cite{BPP:RMP}, but this aspect is of little significance in our
calculation using Eq.~(\ref{eq:tran2}).

With these quantities, $T$ is computed by Monte Carlo integration of
Eq.~(\ref{eq:tran2}).  The integrations also sample
the weight, i. e., the nuclear density.
Figure 2 illustrates the two-particle distribution functions,
$\rho_{pn} ( |{\bf r}|)$ and $\rho_{pp} ( |{\bf r}|)$ for ${}^{16}$O.
For comparison, we also show the distribution functions
obtained using the ${\it A} \Phi$ without the $f$'s.
Our $\rho_{pn}$ and $\rho_{pp}$ have the same shapes as those
by the variational Monte Carlo calculation~\cite{PWP:PRC}
with the peak being at the same location.
But our peak heights are lower than theirs by about 40 \%, and our
$\rho_{pn}$ and $\rho_{pp}$ calculated from ${\it A} \Phi$ alone are
also lower than their mean-field results by a similar amount.
Our one-body density changes little by the application of $f$'s and
remains close to observation, while the one-body density
by the variational calculation changes substantially.

Figures 1 and 2 illustrate our main finding in this letter
that the finite interaction-range
effect reduces the nuclear correlation effect~\cite{Baym:PRC}.
We elaborate on this in the followings.

The values of $T$ are tabulated in Table I, computed by the use of
Eqs.~(\ref{eq:tran2}) and (\ref{eq:tran3}) under various assumptions.
$T$ using Eqs.~(\ref{eq:tran5}) and (\ref{eq:tran6}) is referred to as
{\it eikonal} and is calculated with the one-body density
obtained from Eq.~(\ref{eq:dens}).  We show a)the most naive but popular
calculation using $\sigma_{tot}$ and
b) also with $g(r) = 1 - exp(-{(r/a)}^2)$ ($a=0.7$ fm).
$T$ is increased from a) to b) by about 30 \% and becomes accidentally
rather close to that of our full result in 6), but this
hides the essential physics that we are addressing.

In 1) we show the eikonal result by the use of $\sigma_{inel}$. This
corresponds to the leading-order [Eq.~(\ref{eq:tran4})] calculation
of our full expression [Eq.~(\ref{eq:tran2})]
under the zero-range approximation without correlation,
as it has been noted previously~\cite{Koh:NP}~\cite{Nik:PL}~\cite{Rinat}.
1) is the one that should be
compared with Monte Carlo results of the full expression,
2) --  6) labeled as GMS, {\it Glauber multiple-scattering}.
2) -- 5) examine separately effects of the dynamical and statistical
correlations.  Once the finite-range interaction effect is included,
we find a little variation among $T$'s of 2) -- 6).
6) is our full calculation.
By comparing 1) and 6), we see that the net effect
due to both nuclear correlation and finite-range interaction
is of several \%, above the eikonal $T$ with $\sigma_{inel}$.
This establishes our main finding.
Note that the net effect becomes less in the heavier nucleus.

In the rest of this Letter, we demonstrate that our main finding
is affected little by the choice of the dynamical $ N N $ correlations.
As noted above, the peak heights of our $\rho_{pn}$ and $\rho_{pp}$
differ from those by the variational calculation.
While ours remain close to the observation within about 1 \%,
the RMS radius of the variational ${}^{16}$O density is smaller
than the observed by about 8 \%.
When we artificially enhanced our $f$'s
so as to obtain the variational $\rho_{pn}$ and $\rho_{pp}$, we observed
our RMS radius decreased as much as the variational density did
(though ours had a shallower central depression.)
$T$ computed with the enhanced $f$'s is 0.548 $\pm$ 0.003, which
is {\it smaller} than our full calculation of 6), 0.577 $\pm$ 0.002.
That is, the enhanced $f$'s causes nuclei to be {\it less} transparent.

The reason for the effect is as follows:
At a low nuclear density, $f_c$ around 1 - 2 fm becomes larger than
the asymptotic value of unity, caused by the attractive,
intermediate-range $N N$ interaction~\cite{Wir:Pie}.  The large $f_c$ lifts
the peaks in the two-particle distributions
above the mean field results and increases the one-body density.
Consequently, the knockout proton suffers more final-state interaction.
The variational one-body density differs from observation, and our
wave function is not constructed for the energy minimization.
We conclude that the two-particle distributions are
not reliably known, causing to be $T$ uncertain,
perhaps at about the 5 \% level.

In order to demonstarte this point, we also use a schematic
$f_c$, $1- exp(-{(r/a)}^2)$ ($a=$ 0.7 fm)
by setting all other $f$'s zero.
For this, $\rho_{pn}$ and $\rho_{pp}$ are found to be always below the HF
distributions, the RMS of the density increases from the HF value
of 2.65 to 2.71 fm, and $T$ increases to
0.605 $\pm$ 0.003.  What the correlation does in this case is
merely to enlarge the nuclei and to make them more transparent.
Note that the correlation effect usually discussed in the literature
includes only this short-range repulsive aspect.

In this work, we have examined the reaction that is fully semi-inclusive.
That is, the $E_m$ integration is over the full range, and this permits
the use of closure to simplify our calculation.  A Monte Carlo simulation
is underway to examine how realistic this procedure is by comparing
with the NE18 experiment setup~\cite{MCS}.  Our calculations reported here
are also for simplified kinematics without missing momentum (${\bf p}_m$)
cut.
Our formalism allows ${\bf p}_m$ to be finite, though Eq.~(\ref{eq:tran2})
becomes more complicated, depending on one-particle off-diagonal nuclear
density and the real part of $V_{pN}$.  Furthermore, our formalism
is also applicable to {\it inclusive} $(e,e')$, involving the correlations
differently from the {\it semi-inclusive} $(e,e'p)$ case
considered here.  We will describe a detailed account of
the present work and these considerations in forthcoming publications.

In conclusion, we find that
1) the finite-range $p N$ interaction reduces the effect of
nuclear correlation,
2) the net effect is small and becomes smaller in heavier nuclei, and
3) the precise value of the transparency above the eikonal value
depends sensitively on details of the dynamical correlations in
finite nuclei, which are presently known rather poorly.
\bigskip

We thank S. Pieper and R. Wiringa for the use of
their variational wave functions and for instructive discussions
on nuclear correlations, and T. O'Neil for the use of
his compilation of the $p N$ data.
R.S. and T.S. acknowledge
stimulating and instructive discussions with
members of the Kellogg Radiation Laboratory.
The work was completed while one of us (R.S.) was at the Argonne National
Laboratory.  He thanks R. Wiringa and members of Theory Group for
their kind hospitality.
  This work is supported
by the U.S. DOE at CSUN (DE-FG03-87ER40347) and
by the U.S. NSF at Caltech (PHY91-15574 and PHY94-20470); and
by a Grand-Aid of the Japanese MESC, and Fellowships of JSPS
for Junior Scientists, at U. Tokyo (No. 02640216).

\newpage
\begin{table}
\caption{
Nuclear transparency of closed-shell nuclei at the knockout
proton momentum 4.49 GeV/c in various approximations and models:
With or without $p N$
finite-range interaction (FRI), dynamical correlation (DC),
and statistical correlation (SC) in the eikonal and Glauber
multiple-scattering (GMS) formalism.
The numbers in parentheses denote equations used.
$g$ denotes the use of the simple form of the pair-distribution function
(shown in text)
and POD of a product of one-body densities generated by integarting
the $A$-body density indicated. 5) excludes DC among spectators
(not involving the knockout proton.)
Single-digit values in parentheses are statistical uncertainties
(in the last digits) in Monte Carlo integrations.
}
\bigbreak
\begin{tabular}{cccccccr}
 &FRI & DC & SC & Comments  &${}^{16}{\rm O}$
      & ${}^{40}{\rm Ca}$ \\
\tableline
a)&  &  &  &
${\rm Eikonal}[((\ref{eq:tran5}) ; {\sigma}_{tot}] $
				    & 0.444    & 0.335    \\
b)&  & &  &
${\rm Eikonal} [(\ref{eq:tran6}) ; {\sigma}_{tot}, g ]$
				    & 0.529    & 0.405   \\
 & & & & & & \\

1)&Zero  & No & No &
${\rm Eikonal} [(\ref{eq:tran5}) ; {\sigma}_{inel}] $
				    & 0.560   & 0.446   \\
2)&Finite& No & No &
${\rm GMS} [{\rm POD} ; (|{\it A}\Psi|^2)]$
				    & 0.570(2) &  0.439(2)   \\
3)&Finite& Yes& No &
${\rm GMS} [{\rm POD} ; (\ref{eq:dens})]$
				    & 0.568(2) &   0.442(3)   \\
4)&Finite& No & Yes&
${\rm GMS} [|{\it A}\Psi|^2]$
				    & 0.579(3) &   0.447(3)   \\
5)&Finite& Yes& Yes&
${\rm GMS} [(\ref{eq:dens}); {\rm No spect} ]$
		  & 0.582(2) & 0.450(3) \\
6)&Finite& Yes& Yes&
${\rm GMS} [(\ref{eq:dens}); {\rm Full} ]$
		  & 0.577(2) & 0.447(3) \\
\end{tabular}
\end{table}

\newpage
\begin{figure}
\caption{
$exp(-2 Im \chi ({\bf r}, 0))$ using the spin-independent, local $V_{pp}$
extracted from experimentally determined $p p$ scattering amplitude.
The quantity illustrates attenuation of the knockout proton
as it impinge on a proton located at the origin.
The proton comes from the negative z, parallel to the z-axis.}
\bigskip
\caption{
The two-particle distributions of ${}^{16}$O, $\rho_{pn}$ and $\rho_{pp}$,
calculated using the wave function of Eq.~(\ref{eq:dens}) with and without
pair correlation functions, shown by full curves and open circles,
respectively.}

\end{figure}

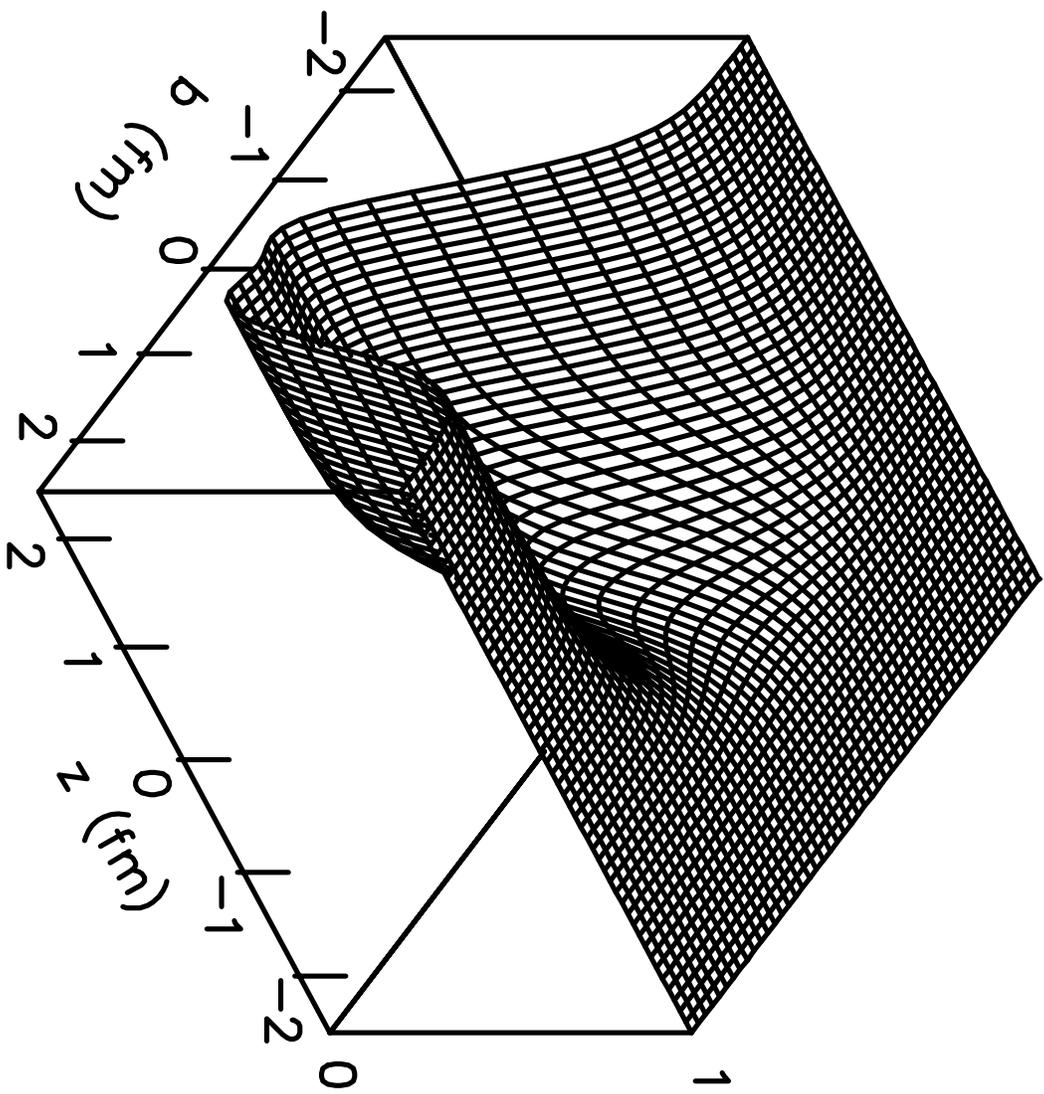
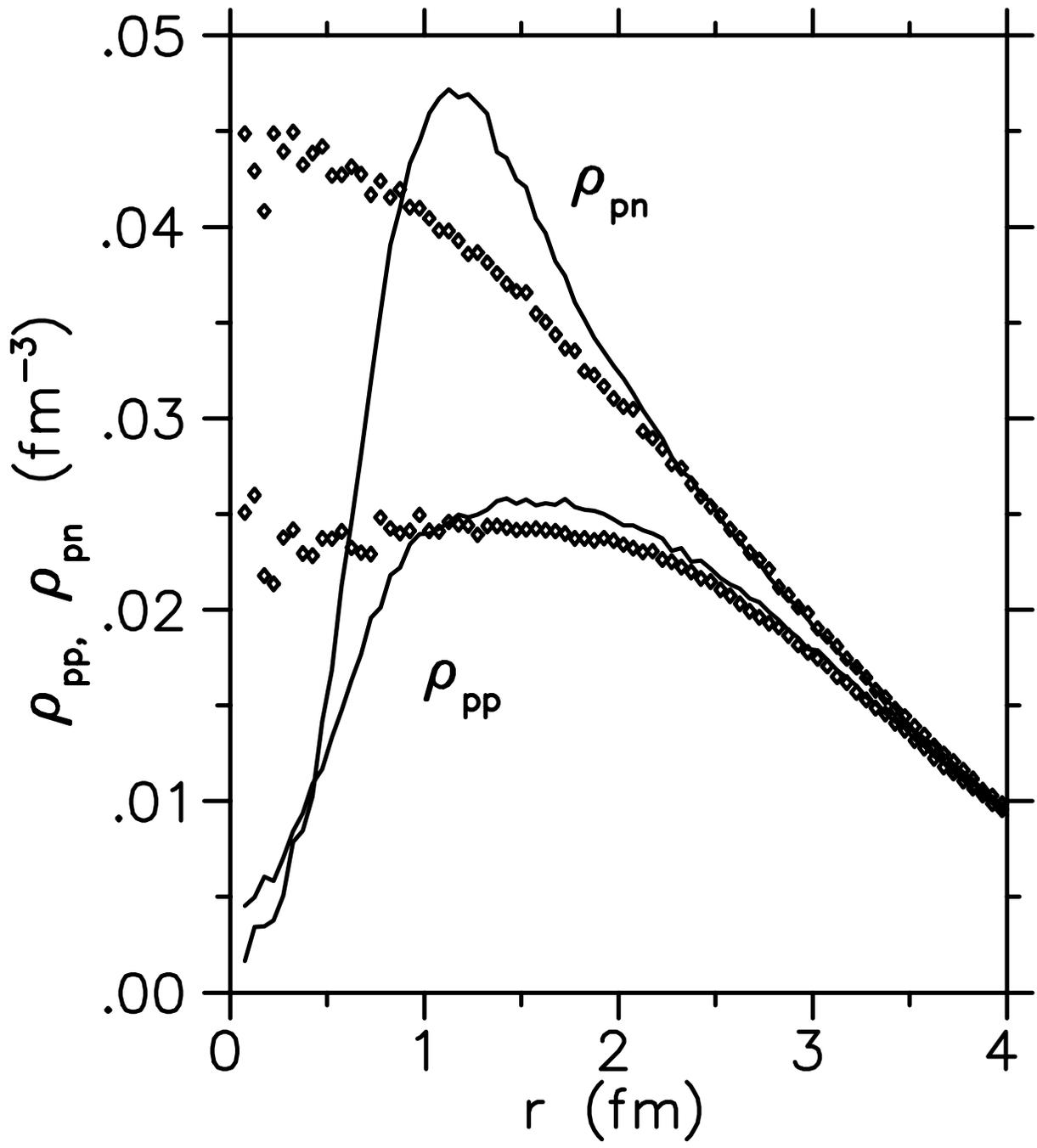
\end{document}